\documentstyle[12pt,epsfig,amssymb,axodraw]{article}
\renewcommand{\arraystretch}{2}
\def\spp{\Sigma_{\scriptscriptstyle ++}}
\def\spm{\Sigma_{\scriptscriptstyle +-}}
\def\smp{\Sigma_{\scriptscriptstyle -+}}
\def\smm{\Sigma_{\scriptscriptstyle --}}
\def\Sp{S_{\scriptscriptstyle +}}
\def\Sm{S_{\scriptscriptstyle -}}
\def\Rp{R_{\scriptscriptstyle +}}
\def\Rm{R_{\scriptscriptstyle -}}
\def\zp{z^\prime}
\def\zplo{z^\prime_{LO}}
\def\alps{\alpha_s}
\def\as{\frac{\alpha_s}{4 \pi}}
\def\xip{\xi^\prime}
\def\hsi{\hat{s}_1}
\def\hti{\hat{t}_1}
\def\hats{\hat{s}}

\def\delp{{\Delta^\prime}}
\def\LN{\ln\left(\frac{\spp+\hsi-\delp}{\spp+\hsi+\delp}\right)}
\def\msbar{${\rm{\overline{MS}}}$}
\def\alps1{${\cal{O}}(\alpha_s^1)$}

\begin{document}
\setlength{\baselineskip}{0.75cm}
\setlength{\parskip}{0.45cm}
\begin{titlepage}
\begin{flushright}
DO-TH 98/05 \linebreak
May 1998
\end{flushright}
\vskip 0.8in
\begin{center}
{\Large\bf Heavy Quark Initiated Contributions to\\ Deep Inelastic
Structure Functions}
\vskip 0.5in
{\large S.\ Kretzer and I.\ Schienbein}
\end{center}
\vskip .3in
\begin{center}
{\large Institut f\"{u}r Physik, Universit\"{a}t Dortmund \\
D-44221 Dortmund, Germany }
\end{center}
\vskip 0.5in
{\large{\underline{Abstract}}}

\noindent
We present ${\cal{O}}(\alpha_s^1 )$ corrections to deep inelastic
scattering amplitudes on massive quarks obtained
within the scheme of Aivazis, Collins,
Olness and Tung (ACOT). 
After identifying the correct subtraction term
the convergence of these contributions towards
the analogous coefficient functions for massless quarks, obtained within
the modified minimal subtraction scheme (${\overline{\rm{MS}}}$), is 
demonstrated. Furthermore, the quantitative relevance of the contributions 
to neutral current (NC)  and charged current (CC) structure functions is 
investigated for several choices of the factorization scale.  
\end{titlepage}
%
\noindent

\section{Introduction}

Leptoproduction of heavy quarks has become a subject of major interest
in QCD 
phenomenology both for experimental and theoretical reasons.
Heavy quark contributions are an important component of measured
neutral current (NC) \cite{h1,zeus} and charged current (CC) \cite{ccfrsf} 
deep inelastic (DI) structure 
functions at lower values of Bjorken-$x$, accessible to present experiments.  
Charm tagging in NC and CC deep inelastic scattering (DIS) offers the 
possibility
to pin down the nucleon's gluon \cite{avogt} and strange sea \cite{cdhsw,ccfrlo,ccfrnlo}
density, respectively, 
both of which
are nearly unconstrained by global fits to inclusive DI data. 
Theoretically it is challenging to understand the production mechanism
of heavy quarks within perturbative QCD. 
The cleanest and most predictive method \cite{grs}
of calculating heavy quark contributions to structure
functions seems to be fixed order perturbation theory (FOPT) where
heavy quarks are produced exclusively by operators built from light
quarks (u,d,s) and gluons (g) and no initial state heavy quark lines show 
up in any Feynman diagram. Heavy quarks produced via FOPT are 
therefore also
called 'extrinsic' since no contractions of heavy quark operators
with the nucleon wavefunction are considered (which in turn would
be characteristic for 'intrinsic' heavy quarks). 
Besides FOPT much effort has been put into formulating
variable flavor number  
schemes (VFNS's) \cite{aot,acot,thoro,mrrs,bmsn} 
which aim at resumming  the quasi-collinear logs [$\ln (Q^2/m^2)$; $Q$ and
$m$ being the virtuality of the mediated gauge boson and the 
heavy quark mass, respectively] arising at
any order in FOPT. All these schemes have in common that 
extrinsic FOPT induces
the boundary condition \cite{ct,bmsn}
$q(x,Q^2=m^2)=0+{\cal{O}}(\alpha_s^2)$ for an intrinsic heavy quark
density, which then undergoes massless renormalization group (RG)
evolution. Apart from their theoretical formulation VFNS's have to be
well understood phenomenologically for a comparison with FOPT and with
heavy quark tagged DI data. 
We will concentrate here on the scheme developed
by Aivazis, Collins, Olness and Tung (ACOT) \cite{aot,acot}. 
In the ACOT scheme full
dependence on the heavy quark mass is kept in graphs containing heavy
quark lines. This gives rise to the above mentioned quasi-collinear logs
as well as to power suppressed terms of ${\cal{O}} [(m^2/Q^2)^k]$. 
While the latter give mass corrections to the massless, dimensionally
regularized, standard coefficient functions (e.g. in the 
${\overline{{\rm{MS}}}}$ scheme), the former are removed by numerical 
subtraction since the collinear region of phase space is already contained in 
the RG evolution of the heavy quark density. 
Up to now explicit expressions
in this scheme exist for DIS on a heavy quark at ${\cal{O}}(\alpha_s^0 )$ 
\cite{aot} as well as for the production of heavy quarks
via virtual boson gluon fusion (GF) at ${\cal{O}}(\alpha_s^1 )$
\ \cite{acot}.
In section 2 we will give expressions which complete the scheme
up to ${\cal{O}}(\alpha_s^1 )$ and calculate DIS on a heavy quark at
first order in the strong coupling, i.e. $B^\ast Q_1 \rightarrow Q_2 g$
(incl. virtual corrections to $B^\ast Q_1 \rightarrow Q_2$) with general 
couplings of the virtual boson $B^\ast$ to the heavy quarks, keeping
all dependence on the masses $m_{1,2}$ of the quarks $Q_{1,2}$. 
It is unclear whether (heavy) quark scattering (QS) and GF at
${\cal{O}}(\alpha_s^1 )$ should be considered on the same level in the
perturbation series. Due to its extrinsic prehistory 
${\rm{QS}}^{(1)} $ (bracketed upper indices count 
powers\footnote{
For the reasons given here we refrain in most cases 
from using the standard terminology of 
'leading' and 'next-to-leading' contributions and count explicit powers of $\alpha_s$.}
of $\alpha_s$)
includes a collinear subgraph of ${\rm{GF}}^{(2)}$,
e.g.\ $\gamma^\ast c\rightarrow c g$ contains the part of 
$\gamma^\ast g \rightarrow c {\bar{c}} g$, where the gluon splits into an
almost on-shell $c {\bar{c}}$ pair.
Therefore QS at ${\cal{O}}(\alpha_s^1 )$ can be considered on the level
of GF at ${\cal{O}}(\alpha_s^{2} )$. On the other hand the standard
counting for light quarks is in powers of $\alpha_s$ 
and heavy quarks should fit in. We therefore suggest that 
the contributions obtained in section 2 should be included in complete
experimental and theoretical NLO-analyses 
which make use of the ACOT scheme. Theoretically the inclusion is required 
for a complete renormalization of the heavy quark density at 
${\cal{O}}(\alpha_s^{1} )$.
However, we leave an ultimate
decision on that point to numerical relevance and present
numerical results in section 3. Not surprisingly they will depend
crucially on the exact process considered (e.g. NC or CC) and the
choice of the factorization scale.
Finally, in section 4 we draw our conclusions.
Appendices A and B outline the calculation of real gluon emission
and virtual corrections, respectively, and some longish formulae 
are presented in Appendix C.

\section{Heavy quark contributions to structure functions}

In this section we will present all contributions to heavy
quark structure functions up to ${\cal{O}}(\alpha_s^1)$. They
are presented analytically  in their fully massive form 
together with the relevant numerical 
subtraction terms which are needed to remove the collinear
divergences in the high $Q^2$ limit. Section 2.1 and 2.3 
contain no new results and are only included for completeness.
In section 2.2 we present our results for the massive analogue
of the massless ${\rm{{\overline{MS}}}}$ coefficient functions
$C_i^{q,{\rm{\overline{MS}}}}$. 
 
\subsection{DIS on a massive quark at ${\cal{O}}(\alpha_s^0)$}

The ${\cal{O}}(\alpha_s^0)$ results for $B^\ast Q_1 \rightarrow Q_2$, 
including mass effects,
have been obtained in \cite{aot} within a helicity basis 
for the hadronic/partonic
structure functions. For completeness and in order to define our normalization
we repeat these results here within 
the standard tensor basis implying the usual structure functions $F_{i=1,2,3}$.
The helicity basis seems to be advantageous since 
in the tensor basis partonic structure functions mix 
to give hadronic structure functions in the presence of masses \cite{aot}. 
However, the mixing matrix is 
diagonal \cite{aot} for the experimental relevant structure functions $F_{i=1,2,3}$ and 
only mixes $F_4$ with $F_5$ which are both suppressed by 
two powers of the lepton
mass. We neglect target (nucleon) mass corrections
which are important at larger values of Bjorken-$x$ \cite{aot} where heavy quark
contributions are of minor importance.

We consider DIS of the virtual Boson $B^\ast$ on the quark $Q_1$ with mass $m_1$
producing the quark $Q_2$ with mass $m_2$. At order ${\cal{O}}(\alpha_s^0)$ 
this proceeds through the parton model diagram in Fig.\ 1 (a).
Finite mass corrections to the massless parton model expressions are
taken into account by adopting the Ansatz given in Eq.\ (4) of \cite{aot}
\begin{equation}
\label{ansatz}
W^{\mu\nu} = \int\ \frac{d\xi}{\xi}\ Q_1(\xi,\mu^2)\ 
{\hat{\omega}}^{\mu\nu}|_{p_1^+=\xi P^+}\ \ \ .
\end{equation}
$W^{\mu\nu}$ is the usual hadronic tensor and ${\hat{\omega}}^{\mu\nu}$ 
is its partonic analogue.
Here as in the following a hat on partonic quantities refers to 
unsubtracted amplitudes, i.e. expressions which still contain mass
singularities in the massless limit.  
$p_1^+$ and $P^+$ are the light-cone momentum components of the incident
quark $Q_1$ and the nucleon, respectively. Generally the '$+$' light-cone component
of a vector $v$ is given by $v^+\equiv (v^0 +v^3)/\sqrt{2}$. 

Contracting the convolution in Eq.\ (\ref{ansatz}) with the projectors in Appendix A
gives the individual hadronic structure functions $F_{i=1,2,3}$. 
In leading order (LO) the latter are given by \cite{aot}
\begin{eqnarray}
F_1^{QS^{(0)}}(x,Q^2)&=&
\frac{\Sp \spp- 2 m_1 m_2 \Sm}{2 \Delta}\ Q_1(\chi,Q^2) 
\nonumber \\
F_2^{QS^{(0)}}(x,Q^2)&=&\frac{\Sp \Delta}{2 Q^2}\ 2 x\ Q_1(\chi,Q^2) 
\nonumber \\
F_3^{QS^{(0)}}(x,Q^2)&=&2 \Rp\ Q_1(\chi,Q^2) 
\label{LO}
\end{eqnarray}
with 
\begin{equation}
\label{pmpm}
\Sigma_{\pm \pm}=Q^2\pm m_2^2\pm m_1^2\ \ \ .
\end{equation}
In Eq.\ (\ref{LO}) we use the shorthand 
$\Delta \equiv \Delta[m_1^2,m_2^2,-Q^2]$ 
, where the usual
triangle function is defined by 
\begin{equation}
\Delta[a,b,c]=\sqrt{a^2+b^2+c^2-2(a b+ b c + c a)}\ \ \ .
\end{equation}
The vector ($V$) and axial vector ($A$) couplings of the 
${\overline{Q}}_2 \gamma_\mu 
(V-A \gamma_5) Q_1$ quark current enter via the following combinations:
\begin{eqnarray}
\label{couplings}
S_{\pm}&=&V{V}^{\prime} \pm A {A}^{\prime}
\nonumber\\
R_{\pm}&=&(V A^{\prime}\pm V^{\prime} A)/2\
\end{eqnarray}
where $V,A \equiv V^{\prime},A^{\prime}$ in the case of pure $B$ scattering
and $V,A \neq V^{\prime},A^{\prime}$ in the case of $B, B^{\prime}$ 
interference (e.g. $\gamma, Z^0$ interference in the standard model). 
The scaling variable $\chi$ generalizes the usual Bjorken-$x$ in the presence
of parton masses and is given by \cite{aot}:
\begin{equation}
\label{chi}
\chi = \frac{x}{2 Q^2}\ \left(\ \spm\ +\ \Delta\ \right)\ \ \ . 
\end{equation}
The mass dependent structure functions in Eq.\ (\ref{LO}) 
motivate the following definitions
\begin{equation}
\label{def}
\left.
\begin{array}{ccc}
{\cal{F}}_1&=&\frac{2 \Delta}{\Sp \spp- 2 m_1 m_2 \Sm}\ F_1 \\
{\cal F}_2&=&\frac{2 Q^2}{\Sp \Delta}\ \frac{1}{2 x}\ F_2 \\
{\cal F}_3&=&\frac{1}{2 \Rp}\ F_3 
\end{array}
\right\}\ =\ Q_1(\chi,Q^2)\ +\ {\cal{O}}(\alpha_s^1)
\end{equation}
such that ${\cal F}_i-{\cal F}_j, \ i,j=1,2,3$, will be finite
of ${\cal O}(\alpha_s)$ in the limit $m_{1,2} \rightarrow 0$.
 
\subsection{DIS on a massive quark at ${\cal{O}}(\alpha_s^1)$}

At ${\cal{O}}(\alpha_s^1)$ contributions from real gluon emission
[Fig.\ 1 (b)] and virtual corrections [Fig.\ 1 (c)] have to be added
to the ${\cal{O}}(\alpha_s^0)$ results of section 2.1.  
The vertex correction with general masses and couplings [Fig.\ 1 (c)]
does to our knowledge not exist in the literature 
and is presented in some detail
in Appendix B. The final result (virtual+real) can 
be cast into the following form:
\begin{eqnarray} 
\label{QS1}
{\hat{{\cal F}}}_{i=1,2,3}^{QS^{(0+1)}} (x,Q^2,\mu^2) &\equiv& 
{\cal F}_i^{QS^{(0)}} (x,\mu^2)+
{\hat{{\cal F}}}_i^{QS^{(1)}} (x,Q^2,\mu^2) \\ \nonumber
= &Q_1(\chi,\mu^2)&
+\frac{\alpha_s(\mu^2)}{2 \pi}\ \int_{\chi}^1\frac{d\xip}{\xip}
\times
\left[Q_1\left(\frac{\chi}{\xip},\mu^2\right) {\hat{H}}_i^q(\xip,m_1,m_2)
\right]\ , \xip\equiv\frac{\chi}{\xi}
\end{eqnarray}
with
\begin{equation}
\label{coefficient}
{\hat{H}}_i^q(\xip,m_1,m_2)=C_F\ \left[(S_i+V_i)\ \delta(1-\xip)\ + 
\frac{1-\xip}{(1-\xip)_{+}}
\  \frac{{\hat{s}}-m_2^2}{8 {\hat{s}}}
\ N_i^{-1}\ \hat{f}_i^Q(\xip)\right]
\end{equation}
where $\hats=(p_1+q)^2$ and the $S_i$, $V_i$, $N_i$ and $\hat{f}_i^Q$ 
are given in Appendix C.
The factorization scale $\mu^2$ will be taken equal to the
renormalization scale throughout.
The '$+$' distribution in Eq.\ (\ref{coefficient}) is a 
remnant of the cancellation of the soft
divergencies from the real and virtual contributions. It is defined as usual:
\begin{equation}
\int_0^1\ d\xip\ f(\xip)\ \left[ g(\xip)\right]_+ \equiv \int_0^1\ 
d\xip\ \left[f(\xip)-f(1)\right]\ g(\xip) \ \ \ .
\end{equation}      
As indicated by the hat on ${\hat{H}}_i^q$, the full massive 
convolution in  Eq.\ (\ref{QS1})
still contains the mass singularity arising from quasi-collinear 
gluon emission from
the initial state quark leg. The latter has to be removed by 
subtraction in
such a way that in the asymptotic limit $Q^2\rightarrow\infty$ 
the well known massless 
${\overline{{\rm{MS}}}}$ expressions are recovered. 
The ${\overline{{\rm{MS}}}}$ limit is
mandatory since all modern parton distributions 
-- and therefore all available heavy
quark densities -- are defined in this particular scheme (or in the DIS scheme 
\cite{aem}, which can be straightforwardly derived from \msbar). 
The correct subtraction
term can be obtained from the following limit
\begin{eqnarray} \nonumber
\lim_{m_1\rightarrow 0} \int_{\chi}^1 \frac{d \xip}{\xip}
Q_1\left(\frac{\chi}{\xip},\mu^2\right) {\hat{H}}_i^q(\xip,m_1,m_2) &=&
\int_{x/\lambda}^1 \frac{d \xip}{\xip}
Q_1\left(\frac{x}{\lambda\xip},\mu^2\right)  
\Bigg\{ H_i^{q,{\overline{{\rm{MS}}}}}(\xip,\mu^2,\lambda)\\ \nonumber &+& C_F 
\ \left[\frac{1+{\xip}^2}{1-\xip}\left(\ln \frac{\mu^2}{m_1^2}-1
-2 \ln(1-\xip)\right)\right]_{+} \Bigg\} \\ 
\label{limit1}
&+& {\cal{O}}\left(\frac{m_1^2}{Q^2}\right)
\end{eqnarray}
where $\lambda = Q^2/(Q^2+m_2^2)$,   
$x/\lambda=\chi|_{m_1=0}$ and 
the $H_i^{q,{\overline{{\rm{MS}}}}}$ 
can be found in Appendix A of \cite{gkr1}. 
Obviously the \msbar\ subtraction term for a 'heavy quark inside a heavy
quark'  is given not only by   
the splitting function $P_{qq}^{(0)} = C_F [(1+{\xi^{\prime}}^2)/(1-\xip)]_+$
times the collinear log $\ln (\mu^2/m_1^2)$
but also comprises a constant term. Herein we agree with Eq.\ (3.15) in
\cite{melnas}\footnote{
We also agree
with the quark initiated coefficient functions in \cite{biol} where
quark masses have been used as regulators}
, where this was first
pointed out in the framework of perturbative fragmentation functions
for heavy quarks. 
We therefore define 
\begin{equation}
{\cal F}_i^{SUB_q} (x,Q^2,\mu^2)=\frac{\alpha_s(\mu^2)}{2 \pi}C_F 
\int_{\chi}^1\frac{d\xip}{\xip}
\left[\frac{1+{\xip}^2}{1-\xip}\left(\ln \frac{\mu^2}{m_1^2}-1
-2 \ln(1-\xip)\right)\right]_{+} Q_1\left(\frac{\chi}{\xip},\mu^2\right)
\label{SUBq}
\end{equation}
such that
\begin{equation}
\label{limit2}
\lim_{m_1\to 0} \left[{\hat{{\cal F}}}_i^{QS^{(1)}} (x,Q^2,\mu^2)
-{\cal F}_i^{SUB_q} (x,Q^2,\mu^2)\right]=
{\cal F}_i^{Q_1^{(1)},\overline{\rm{{MS}}}} (x,Q^2,\mu^2)\ \ \ ,
\end{equation} 
where the superscript $Q_1$ on 
${\cal F}_i^{Q_1^{(1)},\overline{\rm{{MS}}}}$ refers to that
part of the inclusive structure function 
${\cal F}_i$ which is initiated by the 
heavy quark $Q_1$, i.e.\ which is obtained from a convolution with the heavy
quark parton density. 
Note that the limit in Eq.\ (\ref{limit1}) guarantees 
that Eq.\ (\ref{limit2}) is also fulfilled when
$m_1=m_2 \rightarrow 0$ (e.g.\ NC leptoproduction of charm) since 
\begin{equation}
\lim_{m_2 \rightarrow 0} H_i^{q,{\overline{{\rm{MS}}}}}(\xip,\mu^2,\lambda) 
= C_i^{q,{\overline{{\rm{MS}}}}}(\xip,\mu^2)
+ {\cal{O}}\left(\frac{m_2^2}{Q^2}\right) 
\end{equation}
where $C_i^{q,{\overline{{\rm{MS}}}}}$ are the 
standard massless coefficient functions 
in the ${\overline{\rm{{MS}}}}$ scheme, e.g.\ in \cite{aem,fupe}.

\subsubsection{Comparison to existing NC and CC results}

We have performed several cross checks of our results against well known
calculations that exist in the literature \cite{gkr1,biol,aem,fupe,homo}. 
The checks
can be partly inferred from section 2.2. Nevertheless we present here a 
systematic list for the reader's convenience and in order to hint at several
errors which we uncovered in \cite{homo}.

In the charged current case $V=A=1$ our results in Eq.\ (\ref{QS1}) reduce 
in the limit $m_1\rightarrow 0$ to the corresponding
expressions in \cite{biol}, or in
\cite{gkr1}
if the scheme dependent term represented by Eq.\ (\ref{SUBq}) is
taken into account. The latter
agrees with Eq.\ (3.15) in \cite{melnas}. For $m_{1,2}\rightarrow 0$ 
we reproduce the well 
known \msbar\ coefficient functions e.g.\ in \cite{aem,fupe}. 
The vertex correction in Appendix C is implicitly tested because it 
contributes to any of the final results. However, as an independent
cross check the well known QED textbook result can be reproduced 
for $m_1=m_2$, $A=0$.

Initial state parton mass effects in NC DIS at \alps1 have been first 
considered in \cite{homo} within the scenario \cite{incc} of intrinsic
nonperturbative $c\bar{c}$ pairs stemming from fluctuations of the nucleon 
Fock space wavefunction. Although we do not consider such a scenario here we 
note that our results could be easily transferred to corresponding applications
\cite{hsv}. The main difference would be an inclusion of kinematical
target mass effects which are important at larger $x$ \cite{aot}
where a possible nonperturbative charm component is expected \cite{incc} to 
reside. Apart from obvious typos,
we uncovered some errors in \cite{homo} such that
our results differ from those in \cite{homo} by terms that vanish
in the massless limit.
Eq.\ (C.16) in \cite{homo} should be
multiplied by $(1+4 \lambda z^2)$. The typo propagates into 
the final result in 
Eq.\ (51). Furthermore the longitudinal cross section, i.e. Eq.\ (C.17) in
\cite{homo} seems to have been obtained as a residual of $\sigma_R^{(1)}$ and 
the (wrong) $\sigma_R^{(2)}$. Constructing $\sigma_R^{(1)}$ 
from Eqs.\ (C.16), (C.17) via
$\sigma_R^{(1)}=-2\sigma_R^{(L)}+\sigma_R^{(2)}$ reproduces up
to a constant of normalization
the part $\sim \hat{f}_1^Q$ of 
our result for $F_1$ in Eq.\ (\ref{QS1}).
Given the amount of successful independent 
tests of our results we regard the disagreement with \cite{homo} as
a clear evidence that the results in \cite{homo} should be updated
by our calculation. 


\subsection{Gluon fusion contributions at ${\cal{O}}(\alpha_s^1)$}

The gluon fusion contributions to heavy quark structure functions
($B^\ast g \rightarrow {\bar{Q}}_1 Q_2$)
are known for a long time \cite{ggr,gf}
and have been reinterpreted in \cite{acot}
within the helicity basis for structure functions. 
Here we only briefly recall the corresponding formulae in the tensor basis 
for completeness. The GF
component of DI structure functions is given by
\begin{eqnarray} \nonumber
F_{1,3}^{GF}(x,Q^2) &=& \int_{ax}^1\ \frac{d \xip}{\xip}\ g(\xip,\mu^2)\ 
f_{1,3}\left(\frac{x}{\xip},Q^2\right) \\
\label{GF}
F_{2}^{GF}(x,Q^2) &=& \int_{ax}^1\ \frac{d \xip}{\xip}\ \xip g(\xip,\mu^2)\ 
f_{2}\left(\frac{x}{\xip},Q^2\right) 
\end{eqnarray}
where $ax=[1+(m_1+m_2)^2/Q^2]x$ and the $f_i$ can be found for general
masses and couplings in \cite{ggr}. 
The corresponding ${\cal{F}}_i^{GF}$ are obtained from the
$F_i^{GF}$ by using the same normalization factors as
in Eq.\ (\ref{def}).  
Along the lines of \cite{acot} the GF contributions
coexist with the QS contributions which are calculated from the heavy
quark density, which is evolved via the massless RG equations in the
${\rm{{\overline{MS}}}}$ scheme. As already pointed out in section 2.2 the
quasi-collinear log of the fully massive GF term has to be subtracted 
since the corresponding  mass singularities are
resummed to all orders in the massless RG evolution. The subtraction
term for the GF contribution is given by \cite{acot}
\begin{equation}
\label{SUBg}
{\cal F}_i^{SUB_g} (x,Q^2,\mu^2)=
\sum_{k}
\ \frac{\alpha_s(\mu^2)}{2 \pi}
\ \ln \frac{\mu^2}{m_k^2} 
\ \int_{\chi}^1\frac{d\xip}{\xip}\ P_{qg}^{(0)}(\xip)   
\ g\left(\frac{\chi}{\xip},\mu^2\right)\ \ \ ,
\end{equation}  
where $P_{qg}^{(0)}(\xip) = 1/2\ [{\xip}^2+(1-\xip)^2]$. 
Note that Eq.\ (\ref{SUBg}) as well as Eq.\ (\ref{SUBq}) are 
defined relative to the 
${\cal{F}}_i$
in Eq.\ (\ref{def}) and not with respect to the experimental structure
functions $F_i$. The sum in Eq.\ (\ref{SUBg}) runs over the indices of 
the quarks $Q_k$ for which the quasi-collinear logs
are resummed by massless evolution of a heavy quark density, i.e.\
$k=1$, $k=2$ or $k=1,2$.

\subsection{ACOT structure functions at ${\cal{O}}(\alpha_s^1)$}

As already mentioned in the introduction it is not quite clear how
the perturbation series should be arranged for massive quarks, i.e.\
whether the counting is simply in powers of $\alpha_s$ as for light
quarks or whether an intrinsic heavy quark density carries an extra
power of $\alpha_s$ due to its prehistory as an extrinsic particle 
produced by pure GF. We are here interested in the $QS^{(1)}$
component of heavy quark structure functions. Usually the latter is
neglected in the ACOT formalism since it is assumed to be suppressed
by one order of $\alpha_s$ with respect to the GF contribution as
just explained above. It has however been demonstrated in 
\cite{gkr2,kschie} within \msbar\ that
this naive expectation is quantitatively not supported  in the special 
case of semi-inclusive production of charm (dimuon events) in CC DIS. 
We therefore want to investigate the numerical relevance of the
$QS^{(1)}$ contribution to general heavy quark structure functions. In
this article we present results for the fully inclusive case, relevant
for inclusive analyses and fits to inclusive data. We postpone
experimentally more relevant semi-inclusive ($z$-dependent) results
to a future publication \cite{future}. 
Our results at full ${\cal{O}}(\alpha_s^1)$
will be given by
\begin{equation}
\label{complete}
F_i^{(1)} = F_i^{QS^{(0+1)}} + F_i^{GF} - F_i^{SUB_q} - F_i^{SUB_g}
\end{equation}
with  $F_i^{QS^{(0+1)}}$, $F_i^{GF}$, $F_i^{SUB_q}$ and $F_i^{SUB_g}$
given in Eqs.\ (\ref{QS1}), (\ref{GF}), (\ref{SUBq}), and (\ref{SUBg}), 
respectively. 
Furthermore, we will also consider a perturbative expression for
$F_i$ which is constructed along the expectations of the original
formulation of the ACOT scheme, i.e.\ $QS^{(1)}$ is neglected and
therefore $F_i^{SUB_q}$ need not be introduced
\begin{equation}
\label{incomplete}
F_i^{(0)+GF-{SUB_g}} = F_i^{QS^{(0)}} + F_i^{GF} - F_i^{SUB_g}\ \ \ .
\end{equation}

\section{Results for NC and CC structure functions}

In this section we present results which clarify the numerical
relevance of $QS^{(1)}$ contributions to inclusive heavy quark structure
functions in the ACOT scheme. We will restrict ourselves to NC and CC
production of charm since bottom contributions are insignificant to
present DI data. 
Our canonical parton distributions for the NC case
will be CTEQ4M \cite{cteq4}
(Figs.\ 4 and 5 below), which include 'massless heavy partons'
$Q_k$ above the scale $Q^2=m_k^2$. 
Figs.\ 2 and 3, however, have been obtained from the older GRV92 \cite{grv92} 
distributions. 
The newer GRV94 \cite{grv94} parametrizations do not include a 
resummed charm density
since they are constructed exclusively along FOPT.
GRV94 is employed in the CC section. 
The radiative strange sea of GRV94 seems to be closest
to presently available CC charm production data \cite{gkr2}. 
Furthermore, the low
input scale of GRV94 allows for a wide range of variation of the
factorization scale around the presently relevant experimental scales,
which are lower for CC DIS than for NC DIS.
Qualitatively all our results do not depend on the specific
set of parton distributions chosen.   

\subsection{NC structure functions}

For our qualitative analysis we are only considering photon exchange
and we neglect the $Z^0$.
The relevant formulae are all given in section 2 with the following 
identifications:
\begin{eqnarray} \nonumber
Q_{1,2} &\rightarrow& c \\ \nonumber
m_{1,2} \rightarrow m_c &=& 1.6\ (1.5)\ {\rm{GeV\ \ for\ CTEQ4\ (GRV92)}}
\\ \nonumber
V=V^\prime,A=A^\prime &\rightarrow& \frac{2}{3},0
\end{eqnarray}
and we use $\mu^2=Q^2$ if not otherwise noted.
We consider contributions from 
charmed quarks and anti-quarks which are inseparably mixed by
the GF contribution. This means that in Eq.\ (\ref{SUBg}) the sum runs
over $k=1,2$ and the relevant expressions of section 2.1 and 2.2 have
to be doubled [since $c(x,\mu^2)={\bar{c}}(x,\mu^2)$].

First we investigate the importance of finite mass
corrections to the limit in Eq.\ (\ref{limit2}). 
In Fig.\ 2 the difference
${\hat{F}}_2^{QS^{(1)}}-F_2^{SUB_q}$ can be compared to its 
${\rm{{\overline{MS}}}}$ analogue which is 
\begin{equation}
\label{cqmsbar}
F_2^{(c+{\bar{c}})^{(1)},{\rm{\overline{MS}}}} = \frac{4}{9}\ x\  
\frac{\alpha_s(\mu^2)}{2\pi} \left[ (c+{\bar{c}})(\mu^2) \otimes
C_2^{q,{\rm{\overline{MS}}}}\left(\frac{Q^2}{\mu^2}\right)\right] (x,Q^2)
\end{equation}
where $\otimes$ denotes the usual (massless) convolution.
From Fig.\ 2 it is obvious that the 
relative difference between ACOT and ${\rm{{\overline{MS}}}}$ depends crucially
on $x$. It can be large and only slowly convergent to the
asymptotic ${\rm{{\overline{MS}}}}$ limit as can be inferred from Fig.\ 3.
Note that the solid curves in Figs.\ 2, 3 are extremely sensitive to the 
precise definition of the subtraction term in Eq.\ (\ref{SUBq}), 
e.g.\ changing $\chi \rightarrow x$ -- which also removes the collinear 
singularity in the high $Q^2$ limit -- can change the ACOT result by about
a factor of $5$ around $Q^2 \sim  5\ {\rm{GeV}}^2$.\footnote{
The  subtracted gluon contribution GF changes by about a factor 
of $2$ under the same replacement.}  
This is an example of the ambiguities in defining a
variable flavor number scheme which have been formulated in a
systematic manner in \cite{thoro}.    
  
The relative difference  between the subtracted $QS^{(1)}$ contribution
calculated along ACOT and the corresponding ${\rm{{\overline{MS}}}}$
contribution in Eq.\ (\ref{cqmsbar})
appears, however, phenomenologically irrelevant   
if one considers the significance of these contributions 
to the total charm structure function in Fig.\ 4. The complete
${\cal{O}}(\alpha_s^1)$ result (solid line) 
is shown over a wide range of $Q^2$ together with its individual 
contributions from Eq.\ (\ref{complete}). It can be clearly seen that 
the full massive $QS^{(1)}$ contribution is almost completely
canceled by the subtraction term ${SUB_q}$ (Indeed the curves for
$QS^{(1)}$ and ${SUB_q}$ are hardly distinguishable on the scale of
Fig.\ 4). 
The subtracted quark correction is numerically negligible and 
turns out to be indeed suppressed compared to the gluon
initiated contribution, which is also shown in Fig.\ 4. 
Note, however,
that the quark initiated corrections are not unimportant because they
are intrinsically small. Rather the large massive contribution
$QS^{(1)}$ is
perfectly canceled by the subtraction term ${SUB_q}$ provided that 
$\mu^2=Q^2$ 
is chosen.
This is not necessarily the case for different choices of 
$\mu^2$ as we will now 
demonstrate.  
 
In Fig.\ 5 we show the dependence of the complete structure function
and its components on the arbitrary factorization scale
$\mu^2$. 
Apart from the canonical choice $\mu^2=Q^2$ (which was used for all
preceding figures) also different scales have been proposed
\cite{aot2,col}  
like the maximum transverse momentum of the outgoing heavy quark which
is approximately given by $(p_T^{max})^2 \simeq (1/x-1)\ Q^2/4$. For
low values of $x$, where heavy quark structure functions are most
important, the scale $(p_T^{max})^2 \gg  Q^2$. 
The effect of choosing a $\mu^2$ which differs
much from $Q^2$ can be easiest understood for the massless coefficient
functions $C_i^{q,g,{\rm{\overline{MS}}}}$ which contain an
unresummed $\ln (Q^2/\mu^2)$. The latter is of course absent for
$\mu^2=Q^2$ but becomes numerically increasingly important, the more 
$\mu^2$ deviates from $Q^2$. This logarithmic contribution cannot
be neglected since it is the unresummed part of the collinear
divergence which is necessary to define the scale dependence of the
charm density. This expectation is confirmed by Fig.\ 5.
For larger values of $\mu^2$ the subtracted $QS^{(1)}$
contribution is indeed still suppressed relative to the subtracted $GF$
contribution. Nevertheless, its contribution to the total structure function
becomes numerically significant and reaches the $\sim$ 20 \% level around
$(p_T^{max})^2$.   
Note that in this regime the involved formulae of section 2.2 may be
safely approximated by the much simpler convolution in Eq.\
(\ref{cqmsbar}) because they are completely dominated by the universal
collinear logarithm and the finite differences 
${\rm{ACOT}} - {\rm{\overline{MS}}}$
from Figs.\ 2 and 3 become immaterial.
In practice it is therefore always legitimate to approximate the ACOT
results of section 2.2 by their ${\rm{\overline{MS}}}$ analogues
because both are either numerically insubstantial or logarithmically dominated.  
Finally we confirm that
the scale dependence of the
full ${\cal{O}}(\alpha_s^1)$ structure function $F_2^{(1)}$ in
Eq.\ (\ref{complete}) is
larger than the scale dependence of $F_i^{(0)+GF-SUB_g}$ 
in Eq.\ (\ref{incomplete})
which was already pointed out in \cite{cschm}. Nevertheless, the subtracted
$QS^{(1)}$ contribution should be respected for theoretical reasons
whenever $\alpha_s \ln (Q^2/\mu^2)\not\ll 1$.  
 
\subsection{CC structure functions}

Charm production in CC DIS is induced by an $s \rightarrow c$
transition at the $W$-Boson vertex. The strange quark is not really a
heavy quark in the sense of the ACOT formalism, i.e., the production of
strange quarks cannot be calculated reliably at any scale using FOPT because
the strange quark mass is too small. It is nevertheless reasonable to
take into account possible finite $m_s$ effects into perturbative
calculations using ACOT since the subtraction terms remove all long
distance physics from the coefficient functions. Indeed the ACOT
formalism has been used for an experimental analysis of CC charm
production in order to extract the strange sea density of the
nucleon \cite{ccfrnlo}. 
Along the assumptions of ACOT $QS^{(1)}$ contributions have
not been taken into account. This procedure is obviously questionable and has
been shown not to be justified within the ${\rm{\overline{MS}}}$
scheme \cite{gkr2,kschie}. With our results in section 2.2 we can investigate the
importance of quark initiated ${\cal{O}}(\alpha_s^1)$ corrections
within the ACOT scheme for inclusive CC DIS. As already mentioned above,
results for the
experimentally more important case of semi-inclusive ($z$-dependent)
DIS will be presented in a future publication \cite{future}. In the following we
only introduce subtraction terms for collinear divergencies correlated
with the strange mass and treat all logarithms of the charm mass along
FOPT. We do so for two reasons, one theoretical and one experimental:
First, at present experimental scales of CC charm production $\ln
(Q^2/m_c^2)$ terms can be safely treated along FOPT and no
introduction of an a priori unknown charm density is necessary. Second,
the introduction of a subtraction term for the mass singularity of the
charm quark would simultaneously require the inclusion of the 
$c \rightarrow s$ $QS$-transition at the $W$-vertex with no spectator-like 
${\bar{c}}$-quark as in $GF$. This contribution must, however,
be absent when experiments tag on charm in the final state.
CC DIS on massive charm quarks without final-state
charm tagging has been studied
in \cite{dalesio}.    

The numerics of this section can be obtained by the formulae of
section 2 with the following identifications:
\begin{eqnarray} \nonumber
Q_1 \rightarrow s\ \ \ &,& Q_2 \rightarrow c \\ \nonumber
m_{2} \rightarrow m_c &=& 1.5\ {\rm{GeV}}\ {\rm{(GRV94)}}
\\ \nonumber
V=V^\prime,A=A^\prime &\rightarrow& 1,1\ \ \ 
\end{eqnarray}
and the strange mass $m_1=m_s$ will be varied in order to show its
effect on the structure function $F_2^{c}$.

In Fig.\ 6 we show the structure function $F_2^{c}$ and its individual
contributions for two experimental values of $x$ and $Q^2$ \cite{ccfrlo}
under variation of the factorization scale $\mu^2$. Like in
the NC case we show the complete ${\cal{O}}(\alpha_s^1)$ result as
well as $F_2^{(0)+GF-SUB_g}$ where $QS^{(1)}$ has been neglected. 
The thick curves in Fig.\ 6 (a) have been obtained with a regularizing
strange mass of 10 MeV.
They are numerically indistinguishable from the $m_s=0$ \msbar 
\ results
along the lines of \cite{gkr1}. For the thin curves a larger strange mass of
500 MeV has been assumed as an upper limit. Finite mass effects can
therefore be inferred from the difference 
between the thin and the thick curves.
Obviously they are very small for all contributions and can be safely
neglected. For the higher $Q^2$ value of Fig.\ 6 (b) they would be
completely invisible, so we only show the $m_s=$10 MeV 
results ($\equiv$\msbar). 
Since the finite mass corrections within the ACOT scheme turn out to
be negligible as compared to massless \msbar
\ it is not surprising that we confirm the findings of \cite{gkr2,kschie}
concerning the importance of quark initiated corrections.
They are -- in the case of CC production of charm --
{\it{not}} suppressed with respect to gluon
initiated corrections for all reasonable values of the factorization
scale. Only for small choices of $\mu^2 \sim Q^2+m_c^2$ 
can the quark initiated correction be neglected. In this region of $\mu^2$ also
gluon initiated corrections are moderate and Born approximation 
holds within $\sim 10$\% \cite{gkr1}.
For reasons explained in section 3.1
the absolute value of both corrections -- gluon and quark initiated --
become very significant when large factorization scales like
$p_T^{max}$ are chosen. This can be inferred by looking at the region
indicated by the arrow in Fig.\ 6 which marks the scale $\mu=2p_T^{max}$
which was used in \cite{ccfrnlo}. 
Analyses which use ACOT with a
high factorization scale and neglect quark initiated corrections
therefore undershoot the complete \alps1 result by the difference
between the solid and the dot-dashed curve, which can be easily as large as 
$\sim 20$ \%. 
For reasons explained in
the introduction to this section we have used the radiative strange
sea of GRV94. When larger strange seas like CTEQ4 are used the
inclusion of the quark initiated contributions is even more important.

\section{Conclusions}

In this article we have calculated and analysed DIS on massive quarks
at \alps1 within the ACOT scheme for heavy quarks. 
For NC DIS this contribution differs significantly from its massless
\msbar\ analogue for $\mu^2=Q^2$. Both give, however, a very small
contribution to the total charm structure function such that the large
relative difference is phenomenologically immaterial. At higher values of
the factorization scale $\mu^2 \sim (p_T^{max})^2$ the contributions
become significant and their relative difference vanishes. The
$QS^{(1)}$ contribution of section 2.2 can therefore be safely
approximated by its much simpler \msbar\ analogue at any scale. For CC
DIS quark initiated corrections should always be taken into account on
the same level as gluon initiated corrections. Due to the smallness of
the strange quark mass ACOT gives results which are almost identical 
to \msbar.

\section*{Acknowledgements}
We thank E.\ Reya for advice, useful discussions and a careful reading of
the manuscript.
This work has been supported in part by the
'Bundesministerium f\"{u}r Bildung, Wissenschaft, Forschung und
Technologie', Bonn.

\newpage
\setcounter{equation}{0}
\def\theequation{A\arabic{equation}}
\section*{Appendix A: Real Gluon Emission}

We define a partonic tensor
\begin{equation}
\label{ptensor}
{\hat{\omega}}^{\mu\nu} \equiv 
{\overline{\sum_{{\rm{color}}}}} \sum_{\rm{spin}}
\langle Q_2(p_2),g(k)  \left| 
{\overline{Q}}_2 \gamma^\mu (V-A \gamma_5) Q_1
\right| Q_1(p_1) \rangle\ \times \langle \mu \rightarrow \nu \rangle^\ast
\end{equation} 
which can be decomposed into its different tensor components as usual
\begin{equation}
\label{comp}
{\hat{\omega}}^{\mu\nu}=
{-\hat{\omega}_1^Q}\ g^{\mu\nu}+{\hat{\omega}_2^Q}
\ p_1^{\mu} p_1^{\nu}+i{\hat{\omega}_3^Q}
\ \varepsilon_{\alpha\beta}^{\ \ \mu\nu}
p_1^{\alpha} q^{\beta}+{\hat{\omega}_4^Q}
\ q^{\mu} q^{\nu} + {\hat{\omega}_5^Q}
\ (q^{\mu} p_1^{\nu}+q^{\nu} p_1^{\mu})
\ \ \ .
\end{equation}
${\hat{\omega}}_{\mu\nu}$ can be easily calculated from the general Feynman
rules for invariant matrix elements which are customarily expressed 
as functions
of the mandelstam variables ${\hat{s}}\equiv (p_1+q)^2$ and 
${\hat{t}}\equiv (p_1-k)^2$
to which we will refer in the following.  
Projection onto the individual ${\hat{\omega}_{i=1,2,3}^Q}$ in 
Eq.\ (\ref{comp}) is performed for 
nonzero
masses and in $n=4+2\varepsilon$ dimensions with the following operators
\begin{eqnarray} \nonumber
P_1^{\mu\nu} &=& \frac{-1}{2(1+\varepsilon)}\ \Big\{\ g^{\mu\nu}+\big[
\ m_1^2\ q^{\mu} q^{\nu}-Q^2\ p_1^{\mu}\ p_1^{\nu}-
(p_1\cdot q)(q^{\mu}p_1^{\nu}
+p_1^{\mu}q^{\nu})\big] \\ \nonumber
&\times& 4 \Delta^{-2}[m_1^2,{\hat{s}},-Q^2]\ \Big\} \\ \nonumber
P_2^{\mu\nu} &=& 2\ \Big[ -g^{\mu\nu} Q^2+4\ q^{\mu}q^{\nu}
\ \frac{2(1+\varepsilon)(p_1\cdot q)^2
- m_1^2Q^2}{\Delta^2[m_1^2,{\hat{s}},-Q^2]} \\ \nonumber
&+&4(3+2\varepsilon)Q^2\ \frac{
Q^2\ p_1^{\mu}p_1^{\nu}+(p_1\cdot q)(q^{\mu}p_1^{\nu}+p_1^{\mu}q^{\nu})}
{\Delta^2[m_1^2,{\hat{s}},-Q^2]}\Big]
\ \Big\{ (1+\varepsilon)\Delta^2[m_1^2,{\hat{s}},-Q^2]  
\Big\}^{-1}\\ 
P_3^{\mu\nu} &=&\frac{-2 i}{\Delta^2[m_1^2,{\hat{s}},-Q^2]}
\ \varepsilon^{\mu\nu}_{\ \ \lambda\kappa}
\ p_1^{\lambda}\ q^{\kappa}
\label{projectors}
\end{eqnarray}
such that $P_i \cdot {\hat{\omega}} = {\hat{\omega}}_i^Q$. 
The normalization in 
Eqs.\ (\ref{ptensor}), (\ref{comp}) is 
such that real gluon emission contributes $F_i^{R}$ to the hadronic structure 
functions via
\begin{eqnarray} \nonumber
F_1^{R} &=& \frac{1}{8\pi}\ \int_\chi^1 \frac{d \xi}{\xi}\ Q_1(\xi) \int 
d{\rm{\widehat{PS}}}\ {\hat{\omega}}_1^Q \\ \nonumber
F_2^{R} &=& \frac{2 x}{16 \pi} 
\ \int_\chi^1 \frac{d \xi}{\xi}\ \frac{\Delta^2[m_1^2,{\hat{s}},-Q^2]}{2 Q^2}
\ Q_1(\xi) \int 
d{\rm{\widehat{PS}}}\ {\hat{\omega}}_2^Q \\ 
F_3^{R} &=& \frac{1}{8\pi}\ \int_\chi^1 \frac{d \xi}{\xi}
\ \Delta[m_1^2,{\hat{s}},-Q^2]
\ Q_1(\xi) \int 
d{\rm{\widehat{PS}}}\  {\hat{\omega}}_3^Q
\label{hadronic}
\end{eqnarray}
where \cite{gottsch}
\begin{equation}
\label{ps}
\int\ d{\rm{\widehat{PS}}} = \frac{1}{8\pi}\ \frac{{\hat{s}}-m_2^2}{{\hat{s}}}
\ \frac{1}{\Gamma(1+\varepsilon)}
\ \left[ \frac{({\hat{s}}-m_2^2)^2}{4\pi {\hat{s}}}
\right]^\varepsilon\ \int_0^1\ \left[y(1-y)\right]^\varepsilon\ dy
\end{equation}
is the partonic phase space.
In Eq.\ (\ref{ps}) $y$ is related to the partonic centre of 
mass scattering angle
$\theta^\ast$ and the partonic mandelstam variable ${\hat{t}}$ via
\begin{eqnarray} \nonumber
y &\equiv& \frac{1}{2}\ (1+\cos \theta^\ast) \\
  &=& \frac{1}{2\Delta[m_1^2,{\hat{s}},-Q^2]}\ \left[Q^2+m_1^2+{\hat{s}}+
\Delta[m_1^2,{\hat{s}},-Q^2]+
\frac{2{\hat{s}}({\hat{t}}-m_1^2)}{{\hat{s}}-m_2^2}\right]\ \ \ .
\end{eqnarray}
We have chosen dimensional regularization for
the soft gluon poles stemming from ${\hat{s}}\rightarrow m_2^2$ 
which arise from propagators 
in the ${\hat{\omega}}_i$ times phase space factors in $d{\rm{\widehat{PS}}}$.
In Eq.\ (\ref{hadronic}) we use \cite{aem}
\begin{equation}   
(\hats-m_2^2)^{2\varepsilon-1} \sim
\left(1-\frac{\chi}{\xi}\right)^{2\varepsilon-1}=\frac{1}{2\varepsilon}
\ \delta (1-\chi/\xi)+
\frac{1}{(1-\chi/\xi)_+} + {\cal{O}}(\varepsilon)
\label{aem}
\end{equation}
which separates hard gluon emission ($\sim {\hat{f}}_i^Q$) from soft gluon
($S_i$) contributions in Eq.\ (\ref{coefficient}).  
Note that in Eqs.\ (\ref{QS1}), (\ref{coefficient}) the integration variable 
$\xi$, which is implicitly defined
in Eq.\ (\ref{ansatz}), has been changed to $\xip\equiv \chi/\xi$ for an easier
handling of the distributions. For the relation between ${\hat{s}}$ and
$\xip$ see Eq.\ (\ref{s1}) below.

Since all quark masses are kept nonzero, no poles in $y$ 
(collinear singularities)
are contained in the integration volume.
The ${\hat{f}}_i^Q$ which occur in Eq.\ (\ref{coefficient}) and which 
are given below in Eq.\ (\ref{fis}) are therefore 
straightforward integrals of the ${\hat{\omega}}_i^Q$
\begin{equation}
{\hat{f}}_i^Q = \left(g_s^2\ C_F\right)^{-1}
\ \int_0^1\ dy\ {\hat{\omega}}_i^Q
\end{equation} 
and the $S_i$ in Eqs.\ (\ref{coefficient}), (\ref{soft}) pick up the pole 
in Eq.\ (\ref{aem})
\begin{equation}
S_i \sim \frac{1}{\varepsilon}
\ \int_0^1 dy\ \left[y(1-y)\right]^{\varepsilon}
\ \left.\left[{\hat{\omega}}_i^Q (\hats-m_2^2)^2\right]
\right|_{\xi=\chi}\ \ \ ,
\end{equation}
where the proportionality is given by kinematical and phase space factors
which must be kept up to ${\cal{O}}(\varepsilon)$.

The normalization of our hadronic structure functions in Eq.\ (\ref{hadronic}) 
can be clearly inferred from the corresponding LO results in 
Eq.\ (\ref{LO}). Nevertheless, for definiteness we also
give the hadronic differential cross section to which it corresponds
\begin{eqnarray} \nonumber
\frac{d^2 \sigma^{l,{\bar{l}}}}{dx dy} &=& \frac{1}{n_l}
\ \frac{(G_l^{B,B^\prime})^2\ (G_q^{B,B^\prime})^2\ 2 M_N E_{l}}{2 \pi}
\\ &\times& \left[S_{l,+}\ (1-y)F_2 + S_{l,+}\ y^2xF_1 
\pm R_{l,+}\ 2 y(1-\frac{y}{2}) xF_3 \right]\ \ ,
\end{eqnarray} 
where $(G_{l,q}^{B,B^\prime})^2 = 
[ {g_{l,q}^B}^2/(Q^2+M_B^2)\ {g_{l,q}^{B^\prime}}^2/
(Q^2+M_{B^\prime}^2)]^{1/2}$ 
is the effective squared gauge coupling 
-- including the gauge boson propagator --
of the $\gamma_\mu (V-A\gamma_5)$ lepton and quark current, 
respectively, and $n_l$ counts the spin
degrees of freedom of the lepton, e.g.\ $n_l=1,2$ for $l=\nu,e^-$.
The leptonic couplings $S_{l,+},R_{l,+}$ are defined analogous to 
the quark couplings in Eq.\ (\ref{couplings}). As noted below 
Eq.\ (\ref{couplings}),
$B=B^\prime$ for non-interference (pure $B$ scattering). 

\setcounter{equation}{0}
\def\theequation{B\arabic{equation}}
\section*{Appendix B: Vertex Correction}
We have calculated the vertex correction in $n=4+2\varepsilon$ dimensions
at ${\cal{O}}(\alpha_s^1 )$ 
for general masses and couplings
using the Feynman gauge.
The unrenormalized vertex $\Lambda^\mu_0$ [Fig.~1(c.1)] has the structure
\begin{eqnarray}
\Lambda^\mu_0 &=& C_F \as \Gamma(1-\varepsilon) \left(\frac{Q^2}{4 \pi \mu^2}
\right)^{\varepsilon}
\Bigg\{C_{0,-}\ \gamma^\mu L_5 + C_{+}\ \gamma^\mu R_5 
\nonumber\\
&+& C_{1,-}\ m_2\ p_1^\mu\ L_5 
+ C_{1,+}\ m_1\ p_1^\mu\ R_5 
+\ C_{q,-}\ m_2 q^\mu\ L_5 
+ C_{q,+}\ m_1 q^\mu\ R_5 
\Bigg\} 
\end{eqnarray}
with
$L_5=(V-A\ \gamma_5)$, $R_5=(V+A\ \gamma_5)$.
The coefficients read
\begin{eqnarray} 
C_{0,-}&=&\frac{1}{\varepsilon}(-1-\spp I_1)
+\Bigg[\frac{\Delta^2}{2 Q^2}
+\spp \left(1+\ln\left(\frac{Q^2}{\Delta}\right)\right)\Bigg]I_1
\nonumber\\
&+&\frac{1}{2} \ln \left(\frac{Q^2}{m_1^2}\right)
+\frac{1}{2} \ln \left(\frac{Q^2}{m_2^2}\right)
+\frac{m_2^2-m_1^2}{2 Q^2}\ln \left(\frac{m_1^2}{m_2^2}\right)
+\frac{\spp}{\Delta} \\ \nonumber &\times&
\Bigg\{
\frac{1}{2} \ln^2 \left|\frac{\Delta-\spm}{2 Q^2}\right|
+\frac{1}{2} \ln^2 \left|\frac{\Delta-\smp}{2 Q^2}\right|
-\frac{1}{2} \ln^2 \left|\frac{\Delta+\spm}{2 Q^2}\right|
-\frac{1}{2} \ln^2 \left|\frac{\Delta+\smp}{2 Q^2}\right|
\nonumber\\
&-&\ {\rm Li}_2 \left(\frac{\Delta-\spm}{2 \Delta}\right)
-{\rm Li}_2 \left(\frac{\Delta-\smp}{2 \Delta}\right)
+{\rm Li}_2 \left(\frac{\Delta+\spm}{2 \Delta}\right)
+{\rm Li}_2 \left(\frac{\Delta+\smp}{2 \Delta}\right)
\Bigg\} \nonumber
\\ \nonumber
C_{+}&=& 2 m_1 m_2 I_1
\\ \nonumber
C_{1,-}&=&\frac{-1}{Q^2}\left[\spm I_1+\ln \left(\frac{m_1^2}{m_2^2}\right)\right]
\\ \nonumber
C_{1,+}&=&\frac{-1}{Q^2}\left[\smp I_1-\ln \left(\frac{m_1^2}{m_2^2}\right)\right]
\\ \nonumber
C_{q,-}&=&\frac{1}{Q^4}\left[\left(\Delta^2-2 m_1^2 Q^2\right) I_1-2 Q^2 +\spm
\ln \left(\frac{m_1^2}{m_2^2}\right)\right]
\\ 
C_{q,+}&=&\frac{1}{Q^4}\left[\left(-\Delta^2+2 m_2^2 Q^2
-\smp Q^2\right) I_1+2 Q^2 +(\smp+Q^2)
\ln \left(\frac{m_1^2}{m_2^2}\right)\right]
\end{eqnarray}
with
\begin{eqnarray} \nonumber
I_1=\frac{1}{\Delta}\ln\left[\frac{\spp+\Delta}{\spp-\Delta}\right]
\end{eqnarray}
\begin{eqnarray}
I_2&=&I_1 \ln \Delta-\frac{1}{\Delta} \\ \nonumber &\times&
\Bigg\{
\frac{1}{2} \ln^2 \left|\frac{\Delta-\spm}{2 Q^2}\right|
+\frac{1}{2} \ln^2 \left|\frac{\Delta-\smp}{2 Q^2}\right|
-\frac{1}{2} \ln^2 \left|\frac{\Delta+\spm}{2 Q^2}\right|
-\frac{1}{2} \ln^2 \left|\frac{\Delta+\smp}{2 Q^2}\right|
\nonumber\\ 
&-&\ {\rm Li}_2 \left(\frac{\Delta-\spm}{2 \Delta}\right)
-{\rm Li}_2 \left(\frac{\Delta-\smp}{2 \Delta}\right)
+{\rm Li}_2 \left(\frac{\Delta+\spm}{2 \Delta}\right)
+{\rm Li}_2 \left(\frac{\Delta+\smp}{2 \Delta}\right)
\Bigg\}. \nonumber
\end{eqnarray}

The renormalized vertex [Fig.\ 1 (c.1)-(c.3)]
is obtained by wave function renormalization:
\begin{equation}
\Lambda^\mu_R=\gamma^\mu L_5 (Z_1 - 1)+\Lambda^\mu_0+{\cal{O}}(\alpha_s^2)
\end{equation}
where 
$Z_1=\sqrt{Z_2(p_1)Z_2(p_2)}$.
The fermion wave function renormalization constants are defined on mass shell:
\begin{eqnarray}
Z_2(m_i)=1+C_F\as \Gamma(1-\varepsilon)
\left(\frac{m_i^2}{4 \pi \mu^2}\right)^{\varepsilon}
\frac{1}{\varepsilon}[3 - 4 \varepsilon + {\cal{O}}(\varepsilon^2)]
\end{eqnarray}
such that 
\begin{eqnarray}
Z_1= 1+C_F\as \Gamma(1-\varepsilon)
\left(\frac{Q^2}{4 \pi \mu^2}\right)^{\varepsilon}
\left[\frac{3}{\varepsilon}-\frac{3}{2} \ln \left(\frac{Q^2}{m_1^2}\right)
-\frac{3}{2} \ln \left(\frac{Q^2}{m_2^2}\right)-4 \right]\ \ \ .     
\end{eqnarray}

The final result for the renormalized vertex $\Lambda^\mu_R$ reads
\begin{eqnarray}
\Lambda^\mu_R &=& C_F \as \Gamma(1-\varepsilon) 
\left(\frac{Q^2}{4 \pi \mu^2}\right)^{\varepsilon}
\Bigg\{C_{R,-} \gamma^\mu L_5 + C_{+} \gamma^\mu R_5 
\nonumber\\
&+& C_{1,-}\ m_2\ p_1^\mu\ L_5 
+ C_{1,+}\ m_1\ p_1^\mu\ R_5 
+\ C_{q,-}\ m_2\ q^\mu\ L_5 
+ C_{q,+}\ m_1\ q^\mu\ R_5 
\Bigg\} 
\label{ren}
\end{eqnarray}
with $C_{+}$, $C_{1,\pm}$, $C_{q,\pm}$ as given above and
\begin{eqnarray}
C_{R,-}&=&
\frac{1}{\varepsilon}(2-\spp I_1)
+\Bigg[\frac{\Delta^2}{2 Q^2}
+\spp \left(1+\ln\left(\frac{Q^2}{\Delta}\right)\right)\Bigg]I_1 \\ \nonumber
&+&\frac{m_2^2-m_1^2}{2 Q^2}\ln \left(\frac{m_1^2}{m_2^2}\right)
- \ln \left(\frac{Q^2}{m_1^2}\right)
- \ln \left(\frac{Q^2}{m_2^2}\right)
- 4
+\ \frac{\spp}{\Delta} \\ \nonumber &\times&
\Bigg\{
\frac{1}{2} \ln^2 \left|\frac{\Delta-\spm}{2 Q^2}\right|
+\frac{1}{2} \ln^2 \left|\frac{\Delta-\smp}{2 Q^2}\right|
-\frac{1}{2} \ln^2 \left|\frac{\Delta+\spm}{2 Q^2}\right|
-\frac{1}{2} \ln^2 \left|\frac{\Delta+\smp}{2 Q^2}\right|
\nonumber\\
&-&\ {\rm Li}_2 \left(\frac{\Delta-\spm}{2 \Delta}\right)
-{\rm Li}_2 \left(\frac{\Delta-\smp}{2 \Delta}\right)
+{\rm Li}_2 \left(\frac{\Delta+\spm}{2 \Delta}\right)
+{\rm Li}_2 \left(\frac{\Delta+\smp}{2 \Delta}\right)
\Bigg\}. \nonumber
\end{eqnarray}

\setcounter{equation}{0}
\def\theequation{C\arabic{equation}}
\section*{Appendix C: Real and Virtual Contributions to Structure Functions}
The soft real contributions $S_i$ to the coefficient functions in Eq.\
(\ref{coefficient}) are given by
\begin{eqnarray} \nonumber
S_1 &=& \frac{1}{\varepsilon}(-2+\spp I_1)+
2+\frac{\spp}{\Delta} \Big[ \Delta\ I_1 
+ {\rm{Li}}_2\left(\frac{2\Delta}{\Delta-\spp}\right)
- {\rm{Li}}_2\left(\frac{2\Delta}{\Delta+\spp}\right)
\Big] \\ \nonumber
&+&\ln\frac{\Delta^2}{m_2^2 Q^2}\ (-2+\spp I_1) 
\\  S_{2,3}&=&S_1 \label{soft}
\end{eqnarray}
with $I_1$ given in Appendix B and
where $\chi$ is given in Eq.\ (\ref{chi}).
The virtual contributions are derived from the 
renormalized vertex in Eq.\ (\ref{ren})
by using the projectors in Eq.\ (\ref{projectors}): 
\begin{eqnarray} \nonumber
V_1 &=& C_{R,-} + \frac{S_- \spp-2S_+ m_1 m_2}{S_+ \spp-2S_- m_1 m_2}
\ C_+ \\ \nonumber
V_2 &=&  C_{R,-} + \frac{1}{2} \left( m_1^2\ C_{1,+} + m_2^2
\ C_{1,-}\right)+ \frac{S_-}{S_+}\left[C_+ +\frac{m_1
m_2}{2}\left(C_{1,+}+C_{1,-}\right)\right] \\ 
V_3 &=& C_{R,-} + \frac{R_-}{R_+}\ C_+ 
\end{eqnarray}
where the $C$'s are given in Appendix B. Note that the soft poles
($1/\varepsilon$) of $S_i$, $V_i$ cancel in the sum $S_i+V_i$ in 
Eq.\ (\ref{coefficient}) as must be. 

The massive matrix elements $\hat{f}_i^Q(\xip)$ are most conveniently given 
as functions of the mandelstam variable
$\hsi(\xip)\equiv (p_1+q)^2-m_2^2$, i.e. $\hat{f}_i^Q(\xip)
\equiv \hat{f}_i^Q[\hsi(\xip)]$ with
\begin{equation}
\label{s1}
\hsi(\xip)\equiv {\hat{s}}-m_2^2 =
\frac{1-\xip}{2 \xip}[(\Delta-\spm)\xip+\Delta+\spm]\ \ \ .
\end{equation}
From the real graphs of Fig.\ 1 (b) one obtains
\begin{eqnarray} \nonumber
\hat{f}_1^Q(\hsi)&=&
\frac{8}{{\Delta^\prime}^2}\Bigg\{ 
 -\Delta^2 (\Sp \spp -2 m_1 m_2 \Sm) I_{\xip}
+ 2 m_1 m_2 \Sm \Bigg( 
\frac{1}{\hsi} [{\Delta^\prime}^2 + 4 m_2^2 \spm] \\ \nonumber
&+& 2\spm - \smp
+\frac{\spp+\hsi}{2}
+\frac{\hsi+m_2^2}{{\Delta^\prime} \hsi}
\left[{\Delta^\prime}^2 + 2 \spm \spp+ \left(m_2^2+Q^2\right)  
\hsi \right]\ L_{\xip} \Bigg)  
\\ \nonumber
&+& \Sp \Bigg(
\frac{-m_2^2 \spp}{(\hsi+m_2^2)\hsi}(\Delta^2+4 m_2^2 \spm)
-\frac{1}{4 (\hsi+m_2^2)}\Big[3 \spp^2 \smp+4 m_2^2(10 \spp \spm
\\ \nonumber
&-&\spm \smp
- m_1^2 \spp)
+\hsi[-7 \spp \smp+18 \Delta^2-4 m_1^2(7 Q^2-4 m_2^2+ 7 m_1^2)]
\\ \nonumber
&+& 3 \hsi^2[\spm - 2 m_1^2]-\hsi^3\Big]
+\frac{\hsi+m_2^2}{2 {\Delta^\prime}}
\left[\frac{-2}{\hsi} \spp \left(\Delta^2 +2 \spm \spp \right) \right.
\\ \nonumber
&+&\left.\left(4 m_1^2 m_2^2 - 7 \spm \spp \right)
-4\spm\hsi -\hsi^2\right]\ L_{\xip} \Bigg) 
\Bigg\}
\\ \nonumber
\hat{f}_2^Q(\hsi)&=&
\frac{16}{{\Delta^\prime}^4}\Bigg\{ 
 -2 \Delta^4 \Sp I_{\xip}
+ 2 m_1 m_2 \Sm \Bigg(\frac{\hsi+m_2^2}{\Delta^\prime}
\left({\Delta^\prime}^2 - 6 m_1^2 Q^2 \right)
\ L_{\xip} \\ \nonumber
&-&\frac{{\Delta^\prime}^2 (\hsi+\spp)}{2 (\hsi+m_2^2)}
+\left(2{\Delta^\prime}^2 
-3 Q^2 \left(\hsi+\spp \right)\right)\Bigg)  
+ \Sp \Bigg(-2(\Delta^2-6 m_1^1 Q^2)
(\hsi+m_2^2) \\ \nonumber
&-&2\left(m_1^2 + m_2^2\right)\hsi^2
- 9 m_2^2 \spm^2 
+\Delta^2 \left(2\spp-m_2^2 \right)  
+2\hsi \left(2 \Delta^2+\left(m_1^2-5 m_2^2 \right)\spm \right)  
\\ \nonumber
&+& \frac{\left({{\Delta^\prime}}^2 -6 Q^2\left(m_2^2+\hsi \right)\right)\spp 
\left(\hsi+\spp \right)}{2(\hsi+m_2^2)}
-\frac{2 \Delta^2}{\hsi} \left(\Delta^2+2(2 m_2^2+\hsi) \spm \right)
\\ \nonumber
&+& \frac{(\hsi+m_2^2)}
{{\Delta^\prime}}[\frac{-2}{\hsi}\Delta^2(\Delta^2+2\spm \spp)
-2\hsi(\Delta^2-6 m_1^2 Q^2)
\\ \nonumber
&-& ({\Delta^\prime}^2 -18 m_1^2 Q^2) \spp 
- 2 {\Delta^2} \left(\spp+2 \spm \right)]\ L_{\xip}
\Bigg)  
\Bigg\}
\\ \nonumber
\hat{f}_3^Q(\hsi)&=&
\frac{16}{{\Delta^\prime}^2}\Bigg\{ 
- 2 \Delta^2 \Rp I_{\xip}
+2 m_1 m_2 \Rm \left(1-\frac{\smp}{\hsi}+\frac{(\hsi+m_2^2) 
\left(\hsi+\spm \right)}{{\Delta^\prime} \hsi}\ L_{\xip} \right)
\\ \nonumber
&+& {\Rp} \Bigg(\smp-3\spm  - \frac{2}{\hsi} 
\left(\Delta^2 +2 m_2^2 \spm\right) 
- \frac{(\hsi-\smp)(\hsi+\spp) }{2 (\hsi+m_2^2)} 
\\ 
&+& \frac{\hsi+m_2^2}{{\Delta^\prime} \hsi}  
\left[ -\hsi^2 + 4 \left(m_1^2 \smp-\Delta^2\right)  
- 3 \hsi \spm \right]\ L_{\xip}
\Bigg)
\Bigg\}
\label{fis}
\end{eqnarray}
with 
\begin{displaymath}
\displaystyle L_{\xip} \equiv \LN
\end{displaymath}
and
\begin{displaymath}
I_{\xip}=\left(\frac{\hsi+2 m_2^2}{\hsi^2} + 
\frac{\hsi+m_2^2}{\Delta^\prime \hsi^2}\spp\ L_{\xip}\right)\ .  
\end{displaymath}
$\Delta$ is given below Eq.\ (\ref{pmpm}) and
$\Delta^\prime\equiv\Delta[m_1^2,{\hat{s}},-Q^2]$.

Finally, the normalization factors in Eq.\ (\ref{coefficient}) are
\begin{eqnarray}
N_1&=&\frac{\Sp \spp- 2m_1 m_2 \Sm}{2 \Delta},
\qquad N_2=\frac{2 \Sp \Delta}{(\Delta^\prime)^2},\qquad
N_3=\frac{2 \Rp}{\Delta^\prime}\ .
\end{eqnarray}

\newpage
\section*{Figure Captions}
\begin{description}
\item[Fig.\ 1] 
Feynman diagrams for the $QS^{(0)}$ [Fig.\ 1 (a)] and $QS^{(1)}$ 
[Fig.\ 1 (b), (c)]
contributions to ACOT structure functions in Eqs.\ (\ref{LO}) and (\ref{QS1}),
respectively.
\item[Fig.\ 2] 
$x$-dependence of the subtracted $QS^{(1)}$ contribution to the NC charm 
structure function $F_2^c$ \ (solid line). 
$Q^2=\mu^2=10\ {\rm{GeV^2}}$ is fixed.
For comparison the \msbar\ analogue in 
Eq.\ (\ref{cqmsbar}) is shown (dashed line). 
The GRV92 parton distributions have been used. 
\item[Fig.\ 3] 
The same as Fig.\ 2 but varying $Q^2$($=\mu^2$) for fixed $x$.
\item[Fig.\ 4] 
The complete \alps1 neutral current structure function $F_2^c$ and all 
individual 
contributions over
a wide range of $Q^2$, calculated from the CTEQ4M distributions. 
Details of the distinct 
contributions are given in the text.
\item[Fig.\ 5] 
$\mu^2$ dependence of the complete \alps1 NC structure function (CTEQ4M)
in Eq.\ (\ref{complete})
(solid line) and of the structure function in Eq.\ (\ref{incomplete}) 
(dot-dashed line)
where the subtracted $QS^{(1)}$ contribution is neglected. 
Also shown are the different
subtracted \alps1 contributions $GF$ and $QS^{(1)}$. 
\item[Fig.\ 6 (a), (b)] 
The charm production contribution to the charged current structure 
function $F_2$
for a wide range of the factorization scale $\mu^2$ using GRV94. 
The curves are as for the
neutral current case in Fig.\ 5. 
In Fig.\ 6 (a) the thicker curves have been obtained with a 
(purely regularizing) 
strange mass of $10\ {\rm{MeV}}$ which according to Eq.\ (\ref{limit2}) 
(and to the analogous limit for the subtracted $GF$ term \cite{acot})  
numerically reproduces \msbar.
For the thinner curves a strange mass of 
$500\ {\rm{MeV}}$ has been assumed. In Fig.\ 6 (b) all curves correspond to
$m_s=10\ {\rm{MeV}}$ ($\equiv$ \msbar).  
\end{description}

\newpage
\pagestyle{empty}

\textheight 22.0cm
\textwidth 16cm
\oddsidemargin 0.0cm
\evensidemargin 0.0cm
\topmargin 0.0cm
\renewcommand{\arraystretch}{2}
\def\spp{\Sigma_{\scriptscriptstyle ++}}
\def\spm{\Sigma_{\scriptscriptstyle +-}}
\def\smp{\Sigma_{\scriptscriptstyle -+}}
\def\smm{\Sigma_{\scriptscriptstyle --}}
\def\Sp{S_{\scriptscriptstyle +}}
\def\Sm{S_{\scriptscriptstyle -}}
\def\Rp{R_{\scriptscriptstyle +}}
\def\Rm{R_{\scriptscriptstyle -}}
\def\zp{z^\prime}
\def\zplo{z^\prime_{LO}}
\def\alps{\alpha_s}
\def\as{\frac{\alpha_s}{4 \pi}}
\def\xip{\xi^\prime}
\def\hsi{\hat{s}_1}
\def\hti{\hat{t}_1}
\def\delp{{\Delta^\prime}}
\def\LN{\ln\left(\frac{\spp+\hsi-\delp}{\spp+\hsi+\delp}\right)}
\setlength{\baselineskip}{0.75cm}
\setlength{\parskip}{0.45cm}
\begin{center}
\begin{picture}(250,90)(0,-50)
\ArrowLine(50,20)(100,40)
\Vertex(100,40){2.0}
\ArrowLine(100,40)(150,20)
\Photon(100,80)(100,40){3}{6}
\Text(50,10)[]{$p_1$, $m_1$}
\Text(150,10)[]{$p_2$, $m_2$}
\Text(110,60)[]{$q$}
\Text(100,-20)[]{(a)}
\end{picture}
\end{center}
\begin{center}
\begin{picture}(450,90)(0,-50)
\ArrowLine(60,0)(80,20)
\Line(80,20)(100,40)
\Vertex(100,40){2.0}
\ArrowLine(100,40)(170,40)
\Photon(80,80)(100,40){3}{6}
\Gluon(80,20)(125,20){-2}{6}
\Vertex(80,20){2.0}
\Text(60,-10)[]{$p_1$, $m_1$}
\Text(170,30)[]{$p_2$, $m_2$}
\Text(110,60)[]{$q$}
\Text(200,-20)[]{(b)}
\Text(125,10)[]{$k$}
\ArrowLine(260,0)(300,40)
\Vertex(300,40){2.0}
\Line(300,40)(335,40)
\ArrowLine(335,40)(370,40)
\Photon(280,80)(300,40){3}{6}
\Gluon(335,40)(350,10){-2}{6}
\Vertex(335,40){2.0}
\Text(260,-10)[]{$p_1$, $m_1$}
\Text(370,30)[]{$p_2$, $m_2$}
\Text(310,60)[]{$q$}
\Text(350,0)[]{$k$}
\end{picture}
\end{center}
\begin{center}
\begin{picture}(550,90)(0,-50)
\ArrowLine(50,20)(78,31)
\Line(78,31)(100,40)
\Vertex(100,40){2.0}
\Line(100,40)(122,31)
\ArrowLine(122,31)(150,20)
\Photon(100,80)(100,40){3}{6}
\GlueArc(100,40)(25,-157,-23){-3}{6}
\Vertex(78,31){2.0}
\Vertex(122,31){2.0}
\Text(50,5)[]{$p_1$, $m_1$}
\Text(150,5)[]{$p_2$, $m_2$}
\Text(110,60)[]{$q$}
\Text(100,-20)[]{(c1)}
\Line(200,20)(211,25)
\ArrowLine(211,25)(238,35)
\Line(238,35)(250,40)
\Vertex(250,40){2.0}
\Photon(250,80)(250,40){3}{6}
\ArrowLine(250,40)(300,20)
\GlueArc(225,30)(15,-156,24){-2}{10}
\Vertex(211,25){2.0}
\Vertex(238,35){2.0}
\Text(200,5)[]{$p_1$, $m_1$}
\Text(300,5)[]{$p_2$, $m_2$}
\Text(260,60)[]{$q$}
\Text(250,-20)[]{(c2)}
\ArrowLine(350,20)(400,40)
\Vertex(400,40){2.0}
\Photon(400,80)(400,40){3}{6}
\Line(400,40)(411,35)
\ArrowLine(411,35)(438,25)
\Line(438,25)(450,20)
\GlueArc(425,30)(15,-204,-24){-2}{10}
\Vertex(411,35){2.0}
\Vertex(438,25){2.0}
\Text(350,5)[]{$p_1$, $m_1$}
\Text(450,5)[]{$p_2$, $m_2$}
\Text(410,60)[]{$q$}
\Text(400,-20)[]{(c3)}
\end{picture}
\\ {\sl Fig.\ 1}
\end{center}

\newpage
\pagestyle{empty}
\begin{figure}
\vspace*{-1cm}
\hspace*{-1.5cm}
\epsfig{figure=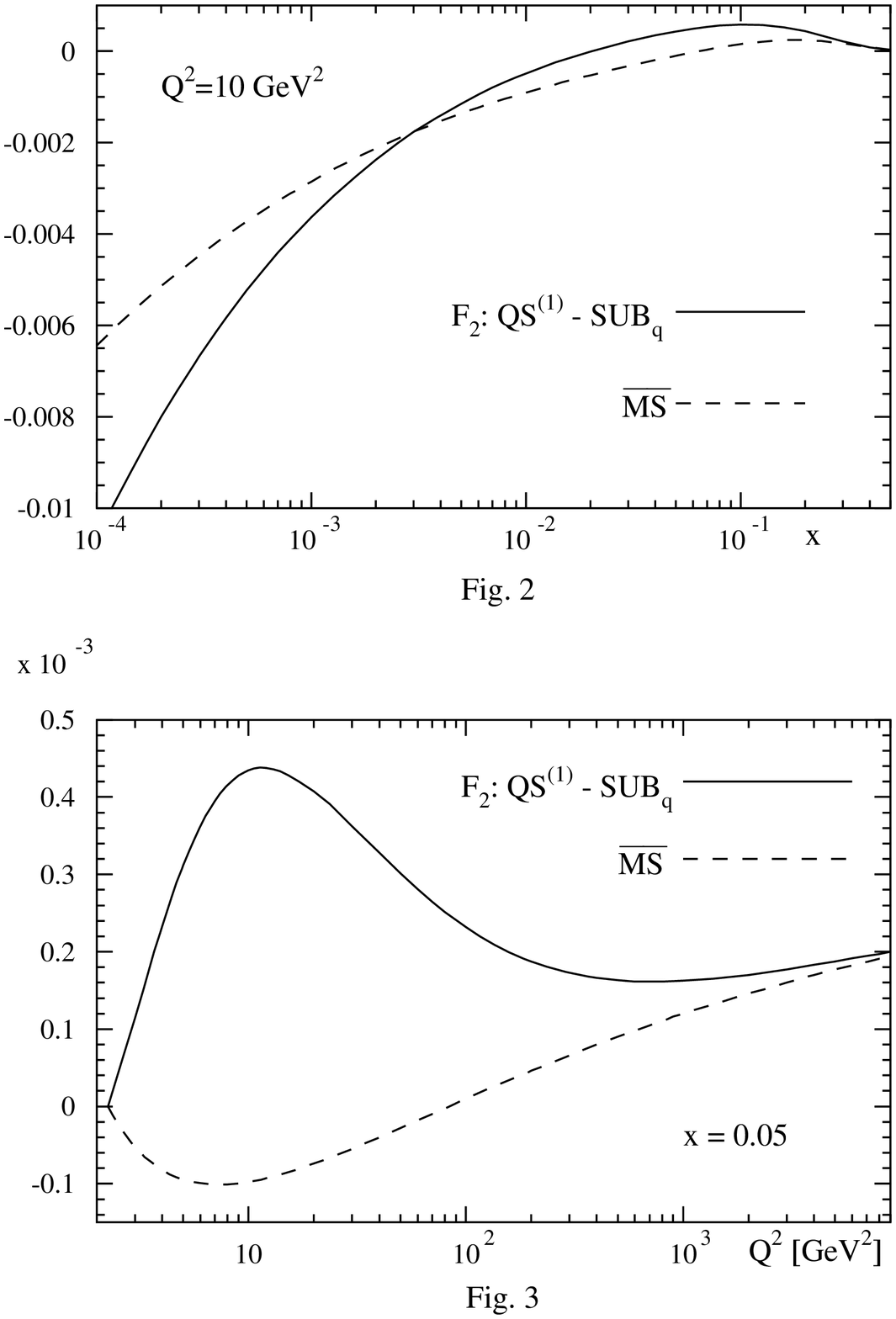,width=20cm}
\end{figure}
\newpage
\begin{figure}
\vspace*{-1.5cm}
\hspace*{-2.5cm}
\epsfig{figure=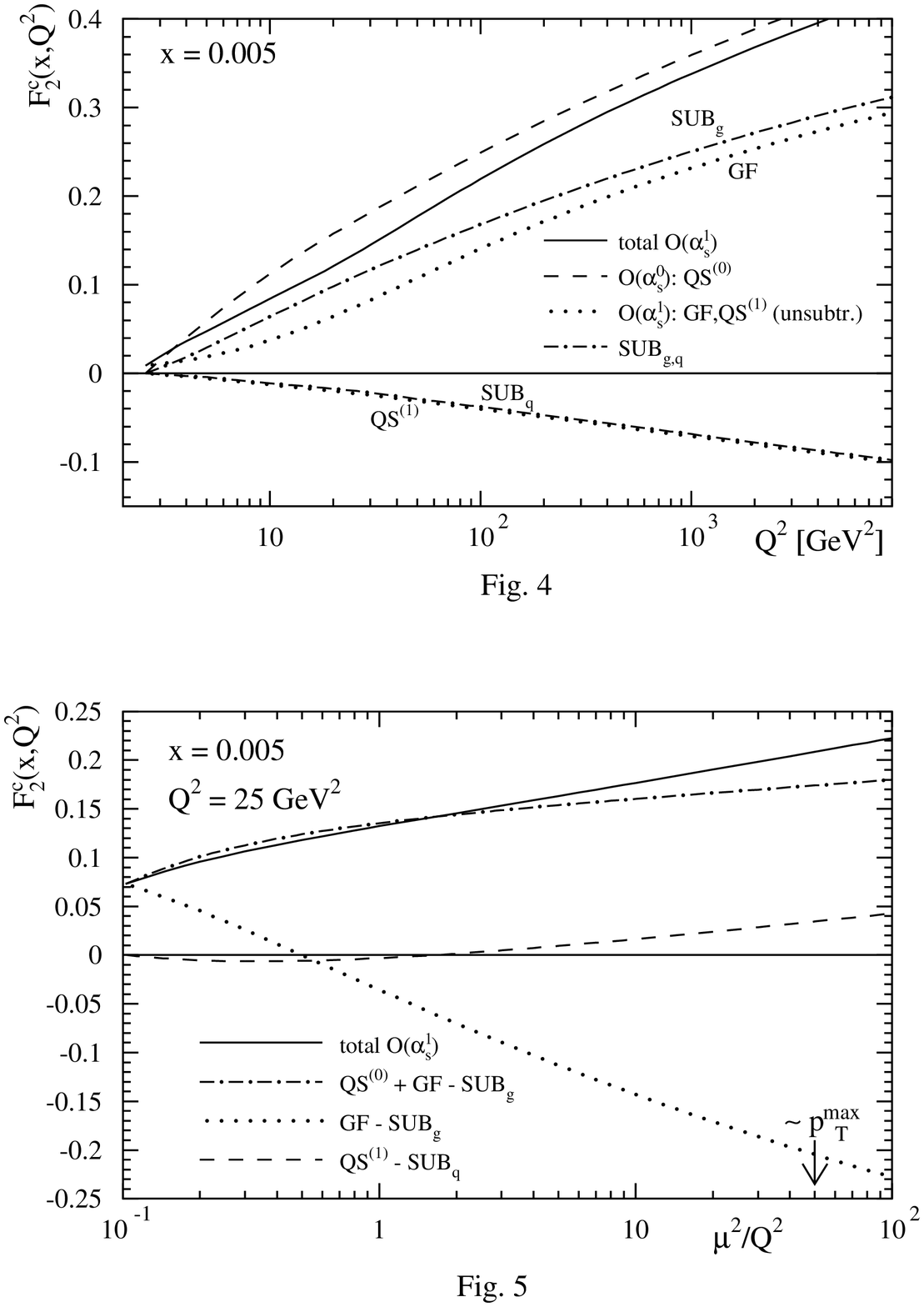,width=20cm}
\end{figure}
\newpage
\begin{figure}
\vspace*{-1.5cm}
\hspace*{-2.5cm}
\epsfig{figure=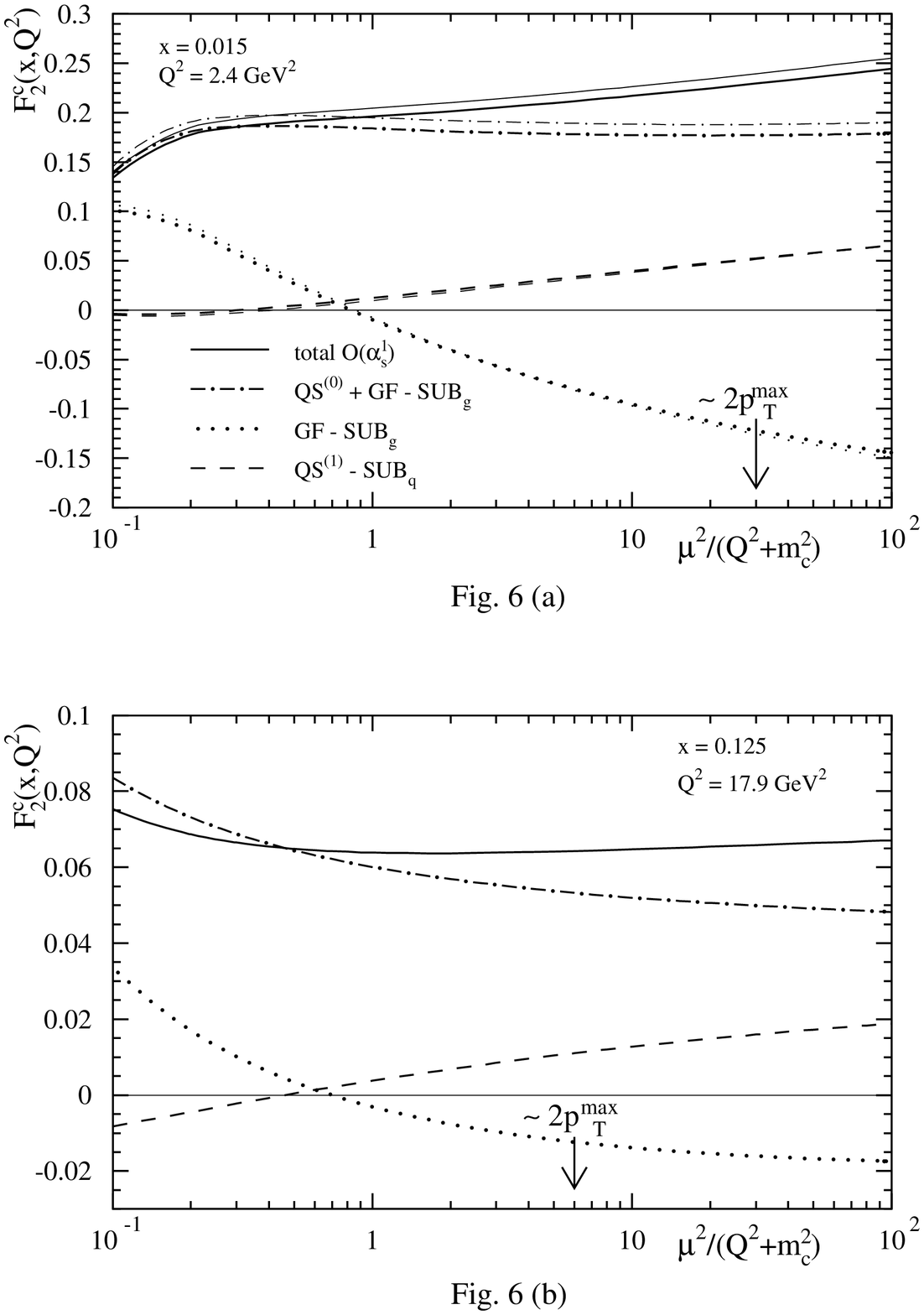,width=20cm}
\end{figure}
\end{document}